\documentclass[aps,prd, twocolumn, showpacs, amssymb, floatfix]{revtex4}
\usepackage{graphicx,amsmath,subfigure}
\begin{document}

\author{Ilya Mandel}
\affiliation{Theoretical Astrophysics, 
	California Institute of Technology, Pasadena, CA 91125}
\email{ilya@caltech.edu}
\date{May 28, 2005}
\title{The geometry of a naked singularity created by standing waves
near a Schwarzschild horizon, and its application to the binary black hole 
problem}
 
\begin{abstract}

The most promising way to compute the gravitational waves emitted by binary 
black holes (BBHs) in their last dozen orbits, where post-Newtonian techniques 
fail, is a {\it quasistationary approximation} introduced by Detweiler and 
being pursued by Price and others.  In this approximation the outgoing
gravitational waves at infinity and downgoing gravitational waves at the holes'
horizons are replaced by standing waves so as to guarantee that the
spacetime has a helical Killing vector field.  Because the horizon generators
will not, in general, be tidally locked to the holes' orbital motion, the 
standing waves will destroy the horizons, converting the black holes into 
naked singularities that resemble black holes down to near the horizon radius.
This paper uses a spherically symmetric, scalar-field model problem
to explore in detail the following BBH issues: (i) The destruction of a horizon 
by the standing waves. (ii) The accuracy with which the resulting naked 
singularity resembles a black hole. (iii) The conversion of the standing-wave
spacetime (with a destroyed horizon) into a spacetime with downgoing waves 
by the addition of a ``radiation-reaction field''. (iv) The accuracy with 
which the resulting downgoing waves agree with the downgoing waves of a 
true black-hole spacetime (with horizon).  The model problem used to study 
these issues consists of a Schwarzschild black hole endowed with spherical 
standing waves of a scalar field, whose wave frequency and near-horizon 
energy density are chosen to match those of the standing gravitational 
waves of the BBH quasistationary approximation.   It is found
that the spacetime metric of the singular, standing-wave spacetime, and 
its radiation-reaction-field-constructed downgoing waves are quite close to
those for a Schwarzschild black hole with downgoing waves --- sufficiently 
close to make the BBH quasistationary approximation look promising for
non-tidally-locked black holes.

\end{abstract}

\pacs{04.25.-g, 04.25.Nx, 04.30.-w}

\maketitle

\section{\label{sec:intro}Introduction and summary}

It is very important, in gravitational astronomy, to have accurate
computations of the gravitational waves from the inspiral of a black 
hole binary \cite{GravRad}.  However, computing these waves is extremely
challenging: for the last $\approx 25$ cycles of inspiral waves, post-Newtonian
approximations fail \cite{Brady}, and numerical relativity can not yet 
evolve the full dynamical equations in this regime.  It appears that the 
best hope for accurately computing the BBH inspiral waves is by a
quasi-stationary approximation \cite{Detweiler, Price}.  In this approximation,
the energy and angular momentum of the binary are conserved by the imposition 
of standing gravitational waves, and the spacetime has a helical Killing vector 
field.  The standing-wave radiation required to keep the orbit stationary is 
computed by demanding that the energy contents of the gravitational waves (GW) 
be minimized \cite{Price}.  

Standing-wave radiation consists of a sum of ingoing and outgoing radiation at 
infinity, and downgoing and upgoing radiation at the black-hole horizons.  
The physical spacetime, with purely outgoing waves at infinity and downgoing 
waves at the horizons, can be recovered from the standing-wave spacetime by 
adding a perturbative radiation-reaction field \cite{private}.  The solution 
for the BBH inspiral consists of a series of quasi-stationary solutions that 
evolve from one to another via energy and angular momentum loss triggered by 
the radiation-reaction field.  The waves measured at a detector can be deduced 
from this sequence of quasistationary solutions.

The black holes comprising the binary are tidally locked if their horizon 
generators are static in the frame co-rotating with the orbit.
In the tidally locked case, the metric perturbations necessary to keep the 
black holes on a stationary orbit are static in the co-rotating frame,
and the black holes can be regarded as having bifurcate Killing horizons
(both a past horizon and a future horizon).  

In reality, the black holes are not tidally locked.  Their mutual tidal forces 
are not strong enough to maintain locking during the inspiral.  In the absence 
of tidal locking, the standing waves of the standing-wave approximation 
destroy the black-hole horizons: the downgoing waves destroy the past horizon 
by building up an infinite energy density at the past horizon, and the 
upgoing waves destroy the future horizon.  Therefore, we expect that forcing 
the orbit to be stationary via the addition of standing gravitational waves 
will strip the Kerr black holes of their horizons and leave naked 
singularities in their place \cite{Wald}.

Despite this radical change in the character of the orbiting bodies, it is 
reasonable to expect that the standing-wave solution will give a quite
accurate approximation to the true physical black-hole spacetimes everywhere
except very near the black-hole horizons.  In order to verify or refute this
expectation, it is necessary to explore the nature of the singularities 
created by the standing gravitational waves and to test how well the physical 
solution with true black holes can be extracted from the standing-wave 
solution with naked singularities.  

As a first step in such an exploration, we consider in this paper a simple 
model problem designed to give insight into the nature of the singularities 
generated by the standing gravitational waves, and the accuracy with which 
the physical, BBH spacetime can be recovered from the standing-wave, 
singularity-endowed spacetime.  

Our model problem is a single, spherically symmetric black hole that is 
converted into a naked singularity by spherical standing waves of a scalar 
field. 

We begin our analysis in Sec.~\ref{sec:sfeq} by describing the mapping 
between the BBH problem, into which we seek insight, and our spherical, 
scalar-field model problem. In particular, we deduce what should be the 
range of scalar-field amplitudes and frequencies in order to mock up the 
gravitational waves of the BBH problem.

Then in Sec.~\ref{sec:sw}, we construct and explore the standing-wave 
spacetime for our spherical model problem.  We initially treat the 
standing-wave scalar field as residing in the unperturbed Schwarzschild
spacetime of the black hole, and we use Regge-Wheeler first-order perturbation 
theory to compute the scalar-energy-induced deviations of the hole's metric
from Schwarzschild.  The metric perturbations consist of a static component 
and a component varying in time at twice the scalar-field frequency 
(see Fig.~\ref{Hpert-fig} below).  The oscillatory component is smaller 
than the static one and higher-order harmonics of both the field and 
the metric are strongly suppressed.  

The static metric perturbation grows divergently as one approaches 
the Schwarzschild horizon --- an obvious indication of the horizon's 
destruction by the standing-wave stress-energy.  To explore the structure 
of the resulting naked singularity, in Sec.~\ref{sec:sw-nonlin} we abandon 
perturbation theory and switch to the fully nonlinear, coupled Einstein 
equations and scalar-field equations.  To simplify the analysis, we focus 
solely on the static part of the singularity's metric; we do this by time 
averaging the scalar stress-energy tensor before inserting it into the 
fully nonlinear Einstein equations.  We solve the resulting equations 
numerically to obtain the spacetime geometry outside the singularity.
The geometry's embedding diagram (Fig.~\ref{emb-fig} below) and the redshift 
seen by a distant observer (Fig.~\ref{red-fig} below) show that the spacetime 
remains nearly Schwarzschild outside the Schwarzschild horizon, but deviates 
strongly from Schwarzschild at $r\approx 2M$ and below.  (Here $M$ is the mass 
of the hole-like singularity and we use geometrized
units $c=G=1$ everywhere in this paper.)  Above $r = 2M$, the standing-wave 
spacetime is very nearly identical to the Schwarzschild spacetime down to 
radii that are well inside the inner edge of the effective potential 
(Fig.~\ref{spacetime-fig}). Below $r = 2M$, radial distance 
changes far more slowly than areal radius; i.e., $g_{rr}$ tends to $0$ as 
$r \rightarrow 0$.  The redshift seen by an external observer rises rapidly 
when the emitter falls inside $r = 2M$.  However, a signal from the 
singularity at $r=0$ may be infinitely redshifted or infinitely blueshifted, 
depending on the choice of scalar field parameters.

In Sec.~\ref{sec:down} we turn to the model spherical spacetime 
that mocks up our desired BBH solution: the spacetime of a Schwarzschild
black hole with downgoing scalar waves.  Not surprisingly, the metric 
perturbations induced by the downgoing scalar-wave energy are those of the 
Vaidya solution of Einstein's equations --- a slowly growing black hole
with a smooth, non-singular future horizon.  This spacetime is well 
approximated, for short time intervals, by the Schwarzschild solution 
with (constant) Schwarzschild mass equal to the instantaneous Vaidya mass.

Finally, in Sec.~\ref{sec:reconstruct} we demonstrate that by adding a 
perturbative radiation-reaction field to the standing-wave solution, a  
downgoing solution to the scalar-wave equation can be recovered.  
We explore the level of agreement between these downgoing waves that live in 
the singularity-endowed standing-wave spacetime and the downgoing
waves in the Schwarzschild approximation to the Vaidya spacetime. 
The agreement (for details see Sec.~\ref{sec:reconstruct} 
and Fig.~\ref{recon-fig} below) is rather good for scalar-wave amplitudes 
and frequencies that mock up the BBH problem --- sufficiently good to 
give optimism that the standing-wave approximation will give accurate 
gravitational waveforms for the final stages of binary-black-hole inspiral.

\section{\label{sec:sfeq}
The mapping between the BBH problem and our model scalar-field problem}

In our exploration of the quasistationary, standing-wave approximation for
black-hole binaries we shall study several spherically symmetric 
spacetimes, each endowed with a standing-wave scalar field.  In 
Sec.~\ref{sec:sw-pert} the spacetime will be Schwarzschild, or Schwarzschild
with first-order gravitational perturbations generated by the scalar-field
stress-energy tensor.  In Sec.~\ref{sec:sw-nonlin} the spacetime will be
that of a naked singularity generated by the coupled, time-averaged 
Einstein-scalar-field equations.  In this section we shall identify the 
parameter regime relevant to gaining insight from these spacetimes into
the binary black hole problem.  

In each of these spherical spacetimes, the scalar field must be a 
solution to the wave equation:
\begin{equation}\label{wave}
\Box \Phi =
\frac{1}{\sqrt{-g}}(\sqrt{-g} g^{\alpha\beta} \Phi_{,\alpha})_{,\beta}=
0 \, ,
\end{equation}
where $g_{\alpha\beta}$ is the spacetime metric with the interval
\begin{equation}\label{genmetric}
ds^2=f(r,t) dt^2 +g(r,t) dr^2 +r^2 (d\theta^2 +\sin^2 \theta d\phi^2) .
\end{equation}
We assume that the scalar field is monochromatic with frequency $\omega$,
and we write it in the form
\begin{equation}\label{Phi}
\Phi=\Re\left ( \frac{\Psi(r) e^{-i\omega t}}{r} \right) \, ,
\end{equation}
where $\Re()$ denotes the real part and the phase was set by the choice
of the zero of time $t$. 

The scalar field $\Phi$ serves as the source of curvature in the Einstein 
equations,
\begin{equation}\label{Einstein}
G_{\alpha\beta}=8\pi T_{\alpha\beta} \, ,
\end{equation}
where the stress-energy tensor depends on the scalar field according to
\begin{equation}\label{T}
T_{\alpha\beta}=\frac{1}{4\pi} \Phi_{,\alpha} \Phi_{,\beta}-
	\frac{1}{8\pi}g_{\alpha\beta} \Phi_{,\mu} \Phi^{,\mu}
\end{equation}
(cf.~Eq.~(20.66) of \cite{MTW} or Eq.~(A.11) of \cite{NF}).

We can re-write equations (\ref{Einstein}) and (\ref{T}) in a simpler form via 
the Ricci tensor:
\begin{equation}\label{Ricci}
R_{\alpha\beta}=2 \Phi_{,\alpha} \Phi_{,\beta} \, .
\end{equation}

Relevant ranges for the scalar-field frequency and amplitude are determined 
by the binary black hole problem we are modeling.  Suppose that the black 
holes in the binary have equal mass $M$, and let $a$ be their radial 
separation.  Since we are interested in the late inspiral, where the 
post-Newtonian  methods fail, the desired range of parameters should 
correspond to $6 \alt a/M \alt 15$ \cite{Brady}.  

The Keplerian orbital frequency of the black holes is
\begin{equation}\label{omegaK} 
\Omega=\frac{1}{M} \sqrt{\frac{2}{(a/M)^3}} \, .
\end{equation}
The gravitational wave frequency is twice the Keplerian frequency, and we 
set our scalar-field frequency equal to the GW frequency:   
\begin{equation}\label{omega}
\omega = 2\Omega = \frac{2}{M} \sqrt{\frac{2}{(a/M)^3}} \, .
\end{equation}

The power going down a black hole due to the orbital motion of its companion 
is approximately
\begin{equation}\label{power}
P_{GW} = \frac{32}{5} M^4 \mu^2 \Omega^6,
\end{equation}
where $\mu$ is the mass of the companion \cite{Poisson, Hua}.  Although the 
calculations in Refs.~\cite{Poisson, Hua} underlying Eq.~(\ref{power}) were 
carried out under the assumption  $\mu \ll M$, we will use Eq.~(\ref{power}) to 
approximate the power for equal mass black holes, $\mu=M$.  This approximation 
is not too worrisome because we are interested in the general features of the 
scalar-field model, which roughly corresponds to the interesting range of BBH 
separations, rather than in the quantitative results for this model.  We 
select the scalar-field amplitude by demanding that its energy density near 
the horizon equal the GW energy density there:
\begin{equation}\label{dEdV}
\frac{dE}{dV} \approx \frac{P_{GW}}{4\pi(2M)^2} \, .
\end{equation}
(In the spirit of this approximate analysis we here ignore the gravitational 
blueshift of the energy.)   By equating this energy density to the value of
$T_{00}$ at the horizon, computed by inserting Eq.~(\ref{Phi}) into 
Eq.~(\ref{T}), we obtain the scalar-field amplitude inside 
the peak of the effective potential:
\begin{equation}\label{amp}
\Psi_{in} = \sqrt{\frac {64}{5}} \left[\frac{1}{(a/M)}\right]^3 M \, .
\end{equation}

Using equations (\ref{omega}) and (\ref{amp}), we can compute the desired
scalar-field frequency and amplitude for the boundaries of the
region of interest:
\begin{subequations}\label{arange}
\begin{equation}\label{a6}
a = 6 M  \Rightarrow \omega \approx 0.19/M, \, \Psi_{in} \approx 0.017 M ;
\end{equation}
\begin{equation}\label{a15}
a = 15 M  \Rightarrow \omega \approx 0.049/M, \, \Psi_{in} \approx 0.0011 M .
\end{equation}
\end{subequations}

\section{\label{sec:sw}Standing-wave scalar field}

We now turn to the standing-wave scalar-field spacetime that mocks up the
spacetimes of the BBH standing-wave approximation.  The metric of this 
spacetime has the form of Eq.~(\ref{genmetric}) and the standing-wave 
scalar field follows from Eq.~(\ref{Phi}):
\begin{equation}\label{Phistat}
\Phi=\frac{\Psi(r) \cos{\omega t}}{r} \, ,
\end{equation}
where $\Psi(r)$ is now real.  

We shall treat the standing-wave scalar field twice via two different 
simplifying  assumptions.  First, in Sec.~\ref{sec:sw-pert}, we will consider 
the scalar field perturbatively; its wave equation will be that of 
the Schwarzschild spacetime, and its stress-energy will generate first-order
perturbations of the metric away from Schwarzschild.  Then in 
Sec.~\ref{sec:sw-nonlin}, we will consider the fully nonlinear
Einstein-scalar-field spacetime but with the scalar stress-energy averaged over
time to make the metric static.

\subsection{\label{sec:sw-pert}Perturbative standing-wave solution}

\subsubsection{Perturbative formalism for the standing-wave spacetime}

\begin{figure}[ht]
\includegraphics[keepaspectratio=true,width=8.5cm,angle=0]{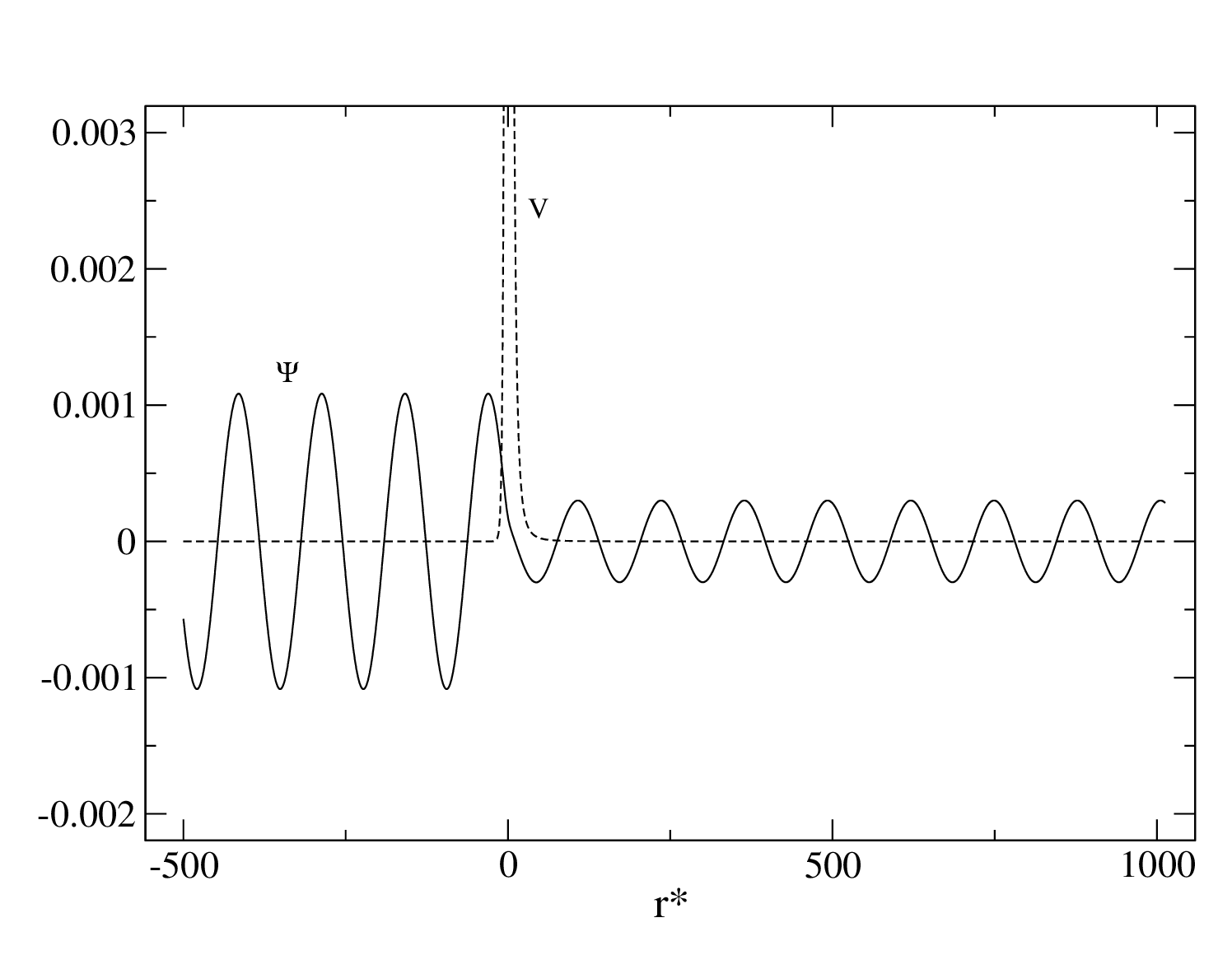}
\caption{The standing-wave scalar field in a Schwarzschild background (solid
curve) and the effective potential (dashed curve) for angular frequency 
$\omega=0.049$.}
\label{Psi-fig}
\end{figure}

In our first approach, the lowest-order solution for the scalar field is 
computed by solving the wave equation (\ref{wave}) in the Schwarzschild 
background with the metric
\begin{eqnarray}\label{metricB}
ds^2 &=& g_{\alpha\beta}^B dx^\alpha dx^\beta \\
\nonumber
	&=& -(1-2/r)dt^2+\frac{1}{1-2/r}dr^2+r^2d\Omega^2 \, ,
\end{eqnarray}
where we rescale so that $M=1$.
The wave equation simplifies as follows (cf.~Eq.~(32.27b) of \cite{MTW}):
\begin{equation}\label{wavepert}
\frac{d^2 \Psi}{{d r^*}^2}=\left[-\omega^2+(1-2/r)\frac{2}{r^3}\right] \Psi \, ,
\end{equation}
where $r^*$ is the Regge-Wheeler tortoise coordinate \cite{RW}, 
\begin{equation}\label{rRW}
r^*=r+2 \ln{(r/2-1)} \, .
\end{equation}
Because $\omega^2$ dominates the right hand side of Eq.~(\ref{wavepert}) 
both far from the horizon ($r \gg 2$) and very near the horizon, the 
scalar field will oscillate with a nearly constant frequency $\omega$ in 
those regions.  In between, where the effective potential 
\begin{equation}\label{V}
V(r^*)=(1-2/r)(2/r^3) ,
\end{equation} 
is significant, there is an intermediate transitional region 
(see Fig.~\ref{Psi-fig}).  (In this paper we mention several times ``the inner
edge of the peak of the effective potential''; we define this inner edge to be
the radius at which the effective potential drops to one percent of 
its maximum value at the peak.)

Since we are approaching the problem perturbatively, we are interested in 
some  small metric perturbation $h_{\alpha\beta}$ on top of the 
background metric $g^B_{\alpha\beta}$ of Eq.~(\ref{metricB}) that would 
yield the curvature corresponding  to the stress-energy tensor of the scalar 
field:  
\begin{equation}\label{metricpert}
g_{\alpha\beta}=g^B_{\alpha\beta} + h_{\alpha\beta} \, .
\end{equation}
Linearizing in $h_{\alpha\beta}$, this metric gives the Ricci tensor
\begin{equation}\label{Riccipert}
R_{\alpha\beta}=R^B_{\alpha\beta} + \frac{1}{2} \left(
	h_{\mu\alpha | \beta}^{\phantom{\mu\alpha | \beta}\mu} + 
	h_{\mu\beta| \alpha}^{\phantom{\mu\beta| \alpha}\mu}
	- h_{\alpha\beta | \mu}^{\phantom{\alpha\beta | \mu}\mu} 
	- h_{| \alpha\beta} \right) \, ,
\end{equation}
where $h=h_{\mu}^{\phantom{\mu}\mu}$ and $_{|}$ represents the covariant 
derivative in the background metric $g^B_{\alpha\beta}$.
For the Schwarzschild background metric, $R^B_{\alpha\beta}=0$.

We are interested only in spherically symmetric perturbations.
A gauge transformation brings additional simplification, so 
$h_{\alpha\beta}$ can be written in the following simple 
Regge-Wheeler form:
\begin{equation}\label{hgennoH1K}
h_{\alpha\beta}= \left( \begin{array}{*{4}{c}} (1-2/r) H_0(t,r) & 0 & 0 & 0 \\
					0 & \frac{H_2(t,r)}{1-2/r} & 0 & 0 \\
					0 & 0 & 0 & 0 \\
					0 & 0 & 0 & 0 
			\end{array}  \right) \, .
\end{equation}
(Compare with Eq.~(13) of \cite{RW} for the case $L=0$.)

We can now substitute $h_{\alpha\beta}$ given by Eq.~(\ref{hgennoH1K}) into 
Eq.~(\ref{Riccipert}) to compute the perturbed Ricci tensor:
\begin{subequations}\label{RiccinoH1K}
\begin{eqnarray}
R_{tt}&=&-\frac{1}{2}\frac{\partial^2}{\partial t^2} H_2 
\\
\nonumber
\lefteqn{
-\frac{r-2}{2 r^3}
	\left[ (2r-1)\frac{\partial}{\partial r} H_0 +
		\frac{\partial}{\partial r}H_2 +
		r(r-2)\frac{\partial^2}{\partial r^2}H_0 \right] \, ;
}
\\
R_{tr}&=&\frac{1}{r} \frac{\partial}{\partial t}H_2 \, ;
\\
R_{rr}&=&\frac{r^2}{2(r-2)^2} \frac{\partial^2}{\partial t^2} H_2 + 
	\frac{1}{2r(r-2)}
\\
\nonumber
\lefteqn{\times
	\left[ 3\frac{\partial}{\partial r} H_0 +
	(2r-3)\frac{\partial}{\partial r} H_2 +
	r(r-2)\frac{\partial^2}{\partial r^2} H_0 \right] \, ;	
}
\\
R_{\theta\theta}&=&H_2 +\frac{r-2}{2} \frac{\partial}{\partial r} H_0 + 
	\frac{r-2}{2} \frac{\partial}{\partial r}H_2 \, .\qquad \qquad \qquad
\end{eqnarray}
\end{subequations}

Inserting expressions (\ref{RiccinoH1K}) for $R_{\alpha\beta}$ into the 
Einstein equations (\ref{Ricci}), one obtains a set of rather complicated 
PDE's containing both spatial and time derivatives to the second order.  
However, we expect that the equations can be further simplified because of 
additional consistency conditions imposed on $\Phi$ by the wave equation 
(\ref{wavepert}).  Indeed, after adding the $R_{tt}$ and $R_{rr}$ equations
with appropriate coefficients to remove the second derivatives in both
$t$ and $r$, and using $R_{\theta\theta}=0$ to relate $H_0$ to
$H_2$, we obtain the following set of first-order ODE's for $H_0$ and $H_2$:
\begin{subequations}\label{H0H2}
\begin{equation}\label{H2}
\frac{\partial H_2}{\partial r} = -\frac{H_2}{r-2} + 
	\frac{r^3}{(r-2)^2} \Phi_{, t} \Phi_{,t} +
	r \Phi_{, r} \Phi_{, r} \, ;
\end{equation}
\begin{equation}\label{H0}
\frac{\partial H_0}{\partial r} = -\frac{\partial H_2}{\partial r} 
				-\frac{2}{r-2}H_2 \, .
\end{equation}
\end{subequations}
These far simpler equations can be shown to produce no spurious solutions; 
in fact, together with the wave equation (\ref{wavepert}),
they are equivalent to the second-order PDE system (\ref{RiccinoH1K}).

\subsubsection{First-order metric perturbations due to the standing-wave 
scalar field}
In the scalar-field ansatz (\ref{Phistat}) we assumed $\Phi \propto 
\cos{\omega t}$.  Therefore, the driving term on the right hand 
side of Eq.~(\ref{H2}) will have static components as well as components 
oscillating in time at the frequency $2\omega$.  Because there is no 
mixing of terms with distinct time signatures in equations (\ref{H0H2}), 
these terms may be treated separately:
\begin{subequations}\label{Hsplit}
\begin{eqnarray}
H_2(t,r) &=& H_2^{stat} (r) + H_2^{cos} (r)  \cos{2\omega t} \, ;
\\
H_0(t,r) &=& H_0^{stat} (r) + H_0^{cos} (r)  \cos{2\omega t} \, .
\end{eqnarray}
\end{subequations}
(There is no $\sin{2\omega t}$ term with our particular choice of the 
scalar-field phase.)

For $r \gg 2$ analytical approximations for $H_0$ and $H_2$ are easy to obtain
because the scalar field is particularly simple there: 
\begin{subequations}\label{Hfar}
\begin{equation}
\Phi \approx (\Psi_0/r) \cos{(\omega r^*)} \cos{(\omega t)} ,
\end{equation}
where $\Psi_0$ is the scalar-field amplitude as $r \rightarrow \infty$.
Inserting this into Eqs.~(\ref{H0H2}), we readily compute, at large $r$:
\begin{eqnarray}
H_2^{stat} (r) &\approx& \frac{1}{2}\omega^2 \Psi_0^2 - \frac{\Psi_0^2}{4r^2} 
	- \frac{\Psi_0^2\cos{2\omega r^*}}{4r^2} \, ;\\
H_2^{cos} (r) &\approx& -\frac{\Psi_0^2}{4r^2} - 
	\frac{\Psi_0^2\cos{2\omega r^*}}{4r^2} \\
	\nonumber
	&&-\frac{\Psi_0^2 \omega \sin{2\omega r^*}}{4r} \, ;\\
H_0^{stat} (r) &\approx& -\omega^2 \Psi_0^2 \ln{r} + 
	\frac{\Psi_0^2\cos{2\omega r^*}}{4r^2}  \, ;\\
H_0^{cos} (r) &\approx& \frac{\Psi_0^2\cos{2\omega r^*}}{4r^2} + 
	\frac{\Psi_0^2 \omega \sin{2\omega r^*}}{4r} \, .
\end{eqnarray}
\end{subequations}
The static components of $H_2$ and $H_0$ are non-vanishing at infinity, 
and $H_0^{stat}$ actually diverges.  This indicates that, due to the energy
contained in the scalar field, the spacetime is not asymptotically flat. 
However, this bad behavior at infinity is an artifact of our model problem 
and is irrelevant to the issues we are studying in this paper.

We can read off from Eqs.~(\ref{Hfar}) the ratios of the 
oscillatory and static components of the metric perturbations at large $r$.  
They are 
\begin{subequations}
\begin{equation}
\left|\frac{H_2^{cos}}{H_2^{stat}}\right| \approx \frac{1}{2\omega r}
\end{equation}
and 
\begin{equation}
\left|\frac{H_0^{cos}}{H_0^{stat}}\right| \approx \frac{1}{4\omega r \ln{r}}\, ;
\end{equation}
\end{subequations}
thus, the static components dominate far from the horizon.

\begin{figure}[ht]
\includegraphics[keepaspectratio=true,width=8.5cm,angle=0]{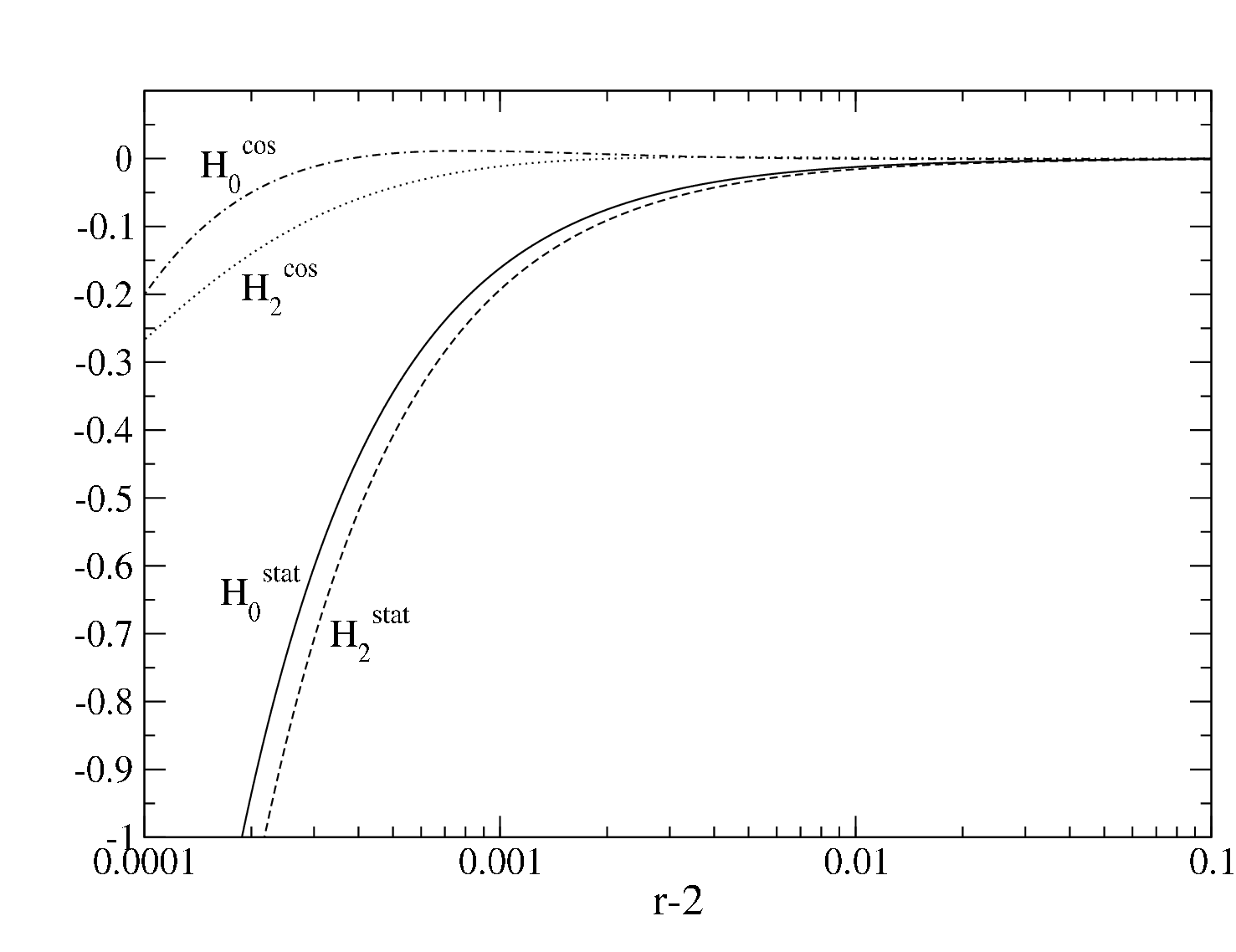}
\caption{Metric perturbations for a standing-wave scalar field in a 
	Schwarzschild background with angular frequency $\omega=0.19$
	and amplitude $\Psi_0=0.015$ far from the black hole,
	corresponding to a binary separation $a \approx 6 M$,
	[Eq.~(\ref{a6})].}
\label{Hpert-fig}
\end{figure}

Equations (\ref{Hfar}) can be used to set initial conditions for the 
metric perturbations at some large $r$, allowing for a numerical solution to 
Eqs.~(\ref{H0H2}) from there down to the horizon, $r=2$.  The resulting 
solution, plotted in  Fig.~\ref{Hpert-fig}, indicates that static components 
continue to dominate near the horizon.

Near the horizon (inside the effective-potential peak), the scalar field has
the form 
\begin{subequations}
\begin{equation}
\Phi \approx (\Psi_{in}/2) \cos{\omega r^*} \cos{\omega t} ,
\end{equation} 
where $\Psi_{in}$ is the scalar-field amplitude as $r \rightarrow 2$.
Inserting this approximation into Eq.~(\ref{H2}) and averaging the right-hand
side leads to the following rough estimate of the magnitude of the 
perturbation near the horizon:
\begin{equation}\label{H2near}
H_2^{stat} \approx \frac{2 \omega^2 \Psi^2 \ln{(r-2)}} {r-2} \, .
\end{equation}
\end{subequations}
Inverting this formula can give a useful estimate of the distance from 
the horizon where the perturbation reaches a particular value; 
the estimate turns out to be accurate to within a factor of two.

Although it appears that the metric perturbation diverges at the expected 
location of the horizon, our perturbative solution is not trustworthy in 
this regime for several reasons in addition to the obvious one of
violating the perturbative assumption $H_0, H_2 \ll 1$:

1. We ignored the {\it back reaction}, i.e., the feedback of the metric 
perturbation into the wave equation.  Using the Schwarzschild metric in place 
of the more accurate perturbed metric in the wave equation, that
is, using the approximate Eq.~(\ref{wavepert}) in place of the exact 
Eq.~(\ref{wave}), is equivalent to an error in the scalar-field
frequency $\Delta \omega / \omega \approx O(H)$, which produces phase offsets 
in the scalar field when the wave equation is integrated numerically.  

2. We linearized the Ricci tensor in the perturbations, neglecting 
higher-order $O(H^2)$ effects. In contrast to the linearized equations 
(\ref{H0H2}) for $H_2$ and $H_0$, the nonlinear perturbative equations are:
\begin{subequations}
\begin{eqnarray}
\label{H2nonlin}
\frac{\partial H_2}{\partial r} &=& -\frac{H_2 (1+H_2)}{r-2}
\\
\nonumber
& &
	+\frac{r^3}{(r-2)^2} \frac{(1+H_2)^2}{1-H_0} (\Phi_{, t})^2 +
	r (1+H_2) (\Phi_{, r})^2 \, ;
\\
\label{H0nonlin}
\frac{\partial H_0}{\partial r} &=& (1-H_0) \left[ 
	-\frac{1}{1+H_2}\frac{\partial H_2}{\partial r} -
	\frac{2 H_2}{r-2} \right] \, .
\end{eqnarray}
\end{subequations}
Linearization introduces local errors of order $H$ into the Einstein 
equations.  However, the errors can build up globally when the 
equations are integrated to obtain a numerical solution.  The errors produced
by linearizing the Ricci tensor (the differences between solutions to the
linearized and nonlinear Einstein equations without back reaction in the 
wave equation) have the same order of magnitude in the parameter range of 
interest as the errors produced by neglecting {\it back reaction} (the 
differences between solutions to the nonlinear Einstein equations 
depending on whether wave equation (\ref{wavepert}) or (\ref{wave}) is used). 

3. We ignored higher harmonics of the scalar field and of the metric 
perturbations that would arise from the back reaction.  However, these higher 
harmonics are suppressed by additional factors of $H \propto \Psi^2$: 
whereas the static and $\cos {2\omega t}$ components of $H$ are quadratic in 
$\Psi$, higher-order harmonics of frequency $2n\omega$ are proportional to 
$\Psi^{2n}$ for $n>1$.

\subsection{\label{sec:sw-nonlin}Time-averaged fully nonlinear 
standing-wave solution}

To explore the behavior of the standing-wave spherical scalar field and the 
spherical metric near and inside the expected location of the horizon, we solve 
the fully nonlinear coupled Einstein-scalar-field equations including full 
back reaction in the wave equation.  To simplify our solution, we average the 
stress-energy tensor in time to remove oscillations of the scalar-field
energy, so that the metric is static.  This is justified by the perturbative 
analysis above, which demonstrates that metric components oscillating in time 
are smaller than static metric components and largely decouple from them.

\subsubsection{Formalism for nonlinear solution with back reaction}
We write the static spherically symmetric metric in the form
\begin{equation}\label{metricstat}
ds^2=-e^{\beta(r)+\alpha(r)}dt^2+e^{\beta(r)-\alpha(r)}dr^2+r^2d\Omega^2 ,
\end{equation}
and we compute the Einstein tensor from this metric in the standard way.  It is
diagonal and its angular components $G_{\hat{\theta}\hat{\theta}}$ and
$G_{\hat{\phi}\hat{\phi}}$ are not particularly interesting because
of spherical symmetry (the angular components of the Einstein equation will 
merely repeat the time and radial components by virtue of the 
contracted Bianchi identities).  The careted subscripts $ _{\hat{\mu}}$ denote
the orthonormal basis associated with the $(t, r, \theta, \phi)$ coordinate
system.  The relevant non-vanishing terms of the Einstein tensor in the
orthonormal basis are:
\begin{subequations}\label{G}
\begin{eqnarray}
G_{\hat{t}\hat{t}}&=&
	e^{\alpha-\beta} (\beta'-\alpha')/r+(1-e^{\alpha-\beta})/r^2 ,
\\
G_{\hat{r}\hat{r}}&=&
	e^{\alpha-\beta} (\beta'+\alpha')/r-(1-e^{\alpha-\beta})/r^2 ,
\end{eqnarray}
\end{subequations}
where $'$ denotes a derivative with respect to $r$, not $r^*$.

Substituting the Einstein tensor (\ref{G}) and the stress-energy tensor 
(\ref{T}) into the Einstein equations (\ref{Einstein}), we obtain: 
\begin{subequations}\label{Ein}
\begin{equation}\label{Ein1}
\alpha'=\frac{1}{r}(e^{\beta-\alpha}-1) \, ,
\end{equation}
\begin{equation}\label{Ein2}
\beta'= r e^{-2 \alpha} (\Phi_{,t})^2 + r (\Phi_{,r})^2 \, .
\end{equation}
\end{subequations}

We can now insert the standing-wave scalar-field ansatz (\ref{Phistat}) and 
time average the right hand side of Eq.~(\ref{Ein2}) over a complete period.
For numerical analysis it will be useful to switch to a logarithmic coordinate
that changes more gradually than $r$ in the vicinity of the Schwarzschild 
horizon.  The following generalization of the Regge-Wheeler tortoise coordinate
$r^*$ proves convenient:
\begin{subequations}\label{Dtot}
\begin{equation}\label{Dr}
\frac{d r}{d r^*} = e^{\alpha} \, .
\end{equation}

In terms of this coordinate, the wave equation (\ref{wave}) simplifies to
\begin{equation}\label{DPsi}
\frac{d^2 \Psi}{d {r^*}^2} = - \omega^2 \Psi + \frac{e^{\alpha}}{r} 
	\frac{d\alpha}{d r^*} \Psi
\end{equation}
and the Einstein equations (\ref{Ein}) with time-averaged 
$(\Phi_{,t})^2$ and $(\Phi_{,r})^2$ become
\begin{equation}\label{Dalpha}
\frac{d \alpha}{d r^*}=\frac{e^{\beta}-e^{\alpha}}{r} \, ,
\end{equation}
\begin{equation}\label{Dbeta}
\frac{d \beta}{d r^*}=
\frac{e^{-\alpha}}{2r}\Biggl[\Psi^2 \omega^2 +\left(\frac{d\Psi}{dr^*}\right)^2
	\Biggr]+
\frac{\Psi^2 e^{\alpha}}{2 r^3} -
\frac{\Psi}{r^2}\frac{d\Psi}{dr^*} \, .
\end{equation}
\end{subequations}

\subsubsection{Singular standing-wave spacetime}
We have solved the coupled equations (\ref{Dtot}) numerically to high
accuracy for values of the scalar-field amplitude and frequency in the range
relevant to the BBH problem [Eqs.~(\ref{arange})].  Our numerical solutions are
very well approximated by analytic formulae that rely on dividing space
$0 < r < \infty$ into three regions. Region I is  ``perturbed 
Schwarzschild'', i.e., the region where the perturbative solution is valid
($r>2$, $H \alt 0.1$).  Region III describes the space very close to 
$r=0$ where the $1/r$ terms diverge.  Finally, the intermediate 
region II extends from the inner boundary of region I to the outer boundary 
of region III.

For sufficiently small amplitudes of the scalar field, the contributions
from the {\it back reaction} (by which we mean the impact of the deviation of
the spacetime from Schwarzschild on the solution to the wave equation) and from
nonlinearity remain small until very close to $r=2$, so that the metric can be 
well approximated by perturbations on top of the Schwarzschild metric.  In
other words, the perturbative solution developed in Sec.~\ref{sec:sw-pert} is
valid throughout region I.  Indeed, for scalar-field amplitudes and frequencies
in the range of interest, the metric perturbations $H_0^{stat}$ and 
$H_2^{stat}$ derived in the previous subsection match the values 
of $H_0$ and $H_2$ corresponding to the complete nonlinear solution with back 
reaction to within $3\%$ for $H \approx 0.01$.

We begin the analysis in region III, where $r \ll 1$, by assuming 
$e^{\beta-\alpha} \ll 1$ as $r \to 0$, which corresponds to $g_{rr} \to 0$ 
at $r=0$.  (This assumption, which can be deduced from the behavior of 
$d\beta / dr^*$ in the transition region, will be shown to be self-consistent; 
more importantly, it is supported by our numerical solutions.)  Then, from
Eq.~(\ref{Dalpha}), $\alpha' \equiv d\alpha/dr \to -1/r$, so $\alpha$ is given 
by
\begin{subequations}\label{tot3}
\begin{equation}\label{alpha3}
\alpha = - \ln r + \alpha_0 \, .
\end{equation}
Here $\alpha_0$ is a constant whose value depends on the amplitude 
and the frequency of the scalar waves; it can be roughly approximated by
\begin{equation}\label{alpha0}
\alpha_0 \sim  \ln{\left(A^2\omega^2\right)}\, ,
\end{equation}
where $A$ is the amplitude of the scalar field near $r=2$.

The wave equation (\ref{DPsi}) becomes
\begin{eqnarray}\label{PPsi3}
\Psi'' &=& - \Psi \omega^2 e^{-2\alpha} - \alpha'(\Psi'-\Psi/r) 
\\
\nonumber
&=&
-\Psi \omega^2 e^{-2\alpha_0} r^2 + 1/r (\Psi'-\Psi/r) \, .
\end{eqnarray}
Since we are interested in the region $r \to 0$, the last term dominates, so
that the approximate solution to Eq.~(\ref{PPsi3}) is
\begin{equation}\label{Psi3}
\Psi = n r + k r \ln r \, ,
\end{equation}
where $n, k$ are constants.

Substituting $\Psi$ and $\alpha$ into Eq.~(\ref{Dbeta}) and selecting 
non-vanishing terms with the highest order in $1/r$, we find that 
$\beta' \to k^2 /(2r)$, so
\begin{equation}\label{beta3}
\beta = \frac{k^2}{2} \ln {r} + \beta_0 \, ,
\end{equation}
where $\beta_0$ is a constant.
Thus, we see that our assumption, $e^{\beta-\alpha} \ll 1$ as $r \to 0$, is 
self-consistent:
\begin{equation}\label{ba3}
\beta-\alpha = \left(\frac{k^2}{2}+1\right) \ln r +\beta_0 -\alpha_0  
	\to -\infty \, \text{as} \, r \to 0 ,
\end{equation}
since the coefficient of $\ln r$ is always positive.

Our numerical solution in region III agrees well with the asymptotic 
analytical solution (\ref{tot3}).  For instance, the value of $k$ obtained 
from matching $\Psi$ to the form of Eq.~(\ref{Psi3}) agrees with the value 
of $k$ obtained from matching $\beta$ to Eq.~(\ref{beta3}) to one part in ten 
thousand.  Of particular interest are the metric components and the
Ricci scalar, whose asymptotics for $r \to 0$ are:
\begin{equation}\label{gtt3}
g_{tt}=-e^{\beta+\alpha}=-e^{\beta_0+\alpha_0} r^{k^2/2 - 1} \, ,
\end{equation}
\begin{equation}\label{grr3}
g_{rr}=e^{\beta-\alpha}=e^{\beta_0-\alpha_0} r^{k^2/2 + 1} \, ,
\end{equation}
and
\begin{equation}\label{Rscalar3}
R=R_{\gamma}^{\gamma}=2\Phi_{,\gamma}\Phi^{,\gamma}=k^2 
	e^{\alpha_0-\beta_0} r^{-3-k^2/2} \, .
\end{equation}
\end{subequations}
The exponent of $r$ in Eq.~(\ref{Rscalar3}) is always negative, so the
Ricci curvature scalar tends to infinity as $r \to 0$, i.e., the radius of 
curvature vanishes at the singularity at $r=0$, as expected. 
The exponent of $r$ in Eq.~(\ref{grr3}) is always positive, so
$g_{rr}$ tends to zero as $r \to 0$ according to a power law.  However,
the sign of the exponent of $r$ in Eq.~(\ref{gtt3}) depends on the value
of $k$, which in turn is a complicated function of the scalar-field 
frequency and amplitude.  For some scalar field parameter values in 
the range relevant to the BBH problem [Sec.~\ref{sec:sfeq}] $k^2/2 > 1$
and $g_{tt}$ vanishes at the singularity; for others, $g_{tt}$ 
is infinite at $r=0$.

The nature of region II, which represents the transition from the 
Schwarzschild-like region I to the singularity of region III, depends strongly 
on the values of $\Psi_0$ and $\omega$.  In Schwarzschild, 
$\alpha=\ln{(1-2/r)}$ tends to $-\infty$ as $r \to 2$, and this is the 
behavior of $\alpha$ in the nearly Schwarzschild region I; meanwhile, in 
region II, as in region III, $\alpha$ is well approximated by
\begin{subequations}\label{tot2}
\begin{equation}\label{alpha2}
\alpha = - \ln r + \alpha_0 \, .
\end{equation}
The outer boundary of region II is located at the transition between these 
two behaviors of $\alpha$, i.e., at the minimum of $\alpha$.  

Substituting the approximation (\ref{alpha2}) for $\alpha$ into the wave 
equation (\ref{DPsi}), we obtain:
\begin{equation}\label{DPsi2}
\frac{d^2\Psi}{d {r^*}^2}=\Psi \bigl(-\omega^2 - \frac{e^{2\alpha_0}}{r^4}\bigr)
\, .
\end{equation}
Thus, the condition for the scalar field to exhibit spatial oscillations at 
an approximately constant amplitude is 
$e^{2\alpha_0} / r^4 \ll \omega^2$.  The location where this condition
begins to be violated forms the inner boundary of region II.  Thus, region II
can be said to be defined by the variation of $\alpha$ according to
Eq.~(\ref{alpha2}) as in region III, and by rapid spatial oscillations of 
the scalar field $d\Psi / dr^* = \omega$ as in region I.

Since $\alpha_0$ will be more negative for smaller amplitudes of the scalar 
field, we see that region II is going to be significant for small $\Psi_0$,
including those in the relevant range of the BBH problem.  For larger values 
of $\Psi_0$, the metric and scalar field will proceed directly from region 
I to region III.

When region II does exist, the amplitude and phase of the scalar field 
[solution of Eq.~(\ref{DPsi2})]
\begin{equation}\label{Psi2}
\Psi(r) = A(r) \cos {\phi(r)}
\end{equation}
will be given by
\begin{equation}\label{A2}
A = A_0 (1 - \frac{e^{2\alpha_0}}{4 r^4 \omega^2} + ...) \, ,
\end{equation}
\begin{equation}\label{phi2}
\dot{\phi}=\omega (1 + \frac{e^{2\alpha_0}}{2 r^4 \omega^2} + ...) \, ,
\end{equation}
to first order in $e^{2\alpha_0}/(r^4 \omega^2)$.

Substituting expressions (\ref{tot2}) for $\alpha$ and $\Psi$ into the 
differential equation for $\beta$, Eq.~(\ref{Dbeta}), we find that the dominant 
term is the first one,
$d\beta / dr^* \to (1/2) e^{-\alpha_0} A^2 \omega^2$, so in region II
$\beta$ is approximately
\begin{eqnarray}
\nonumber
\beta &=& \frac{1}{2} e^{-\alpha_0} A^2 \omega^2 r^* + \text{const} 
\\
\label{beta2}
	&=& \frac{1}{4} e^{-2\alpha_0} r^2 A^2 \omega^2 +\text{const} \, ,
\end{eqnarray}
where the last equality comes from the integral of equation (\ref{Dr}),
$r^* = e^{-\alpha_0} r^2 / 2 + \text{const}$.
\end{subequations}

Embedding diagrams and redshifts may provide the best pictorial insight into 
our full time-averaged standing-wave scalar-field solution, including all of
regions I, II and III.  

\begin{figure}[ht]
\includegraphics[keepaspectratio=true,width=8.5cm,angle=0]{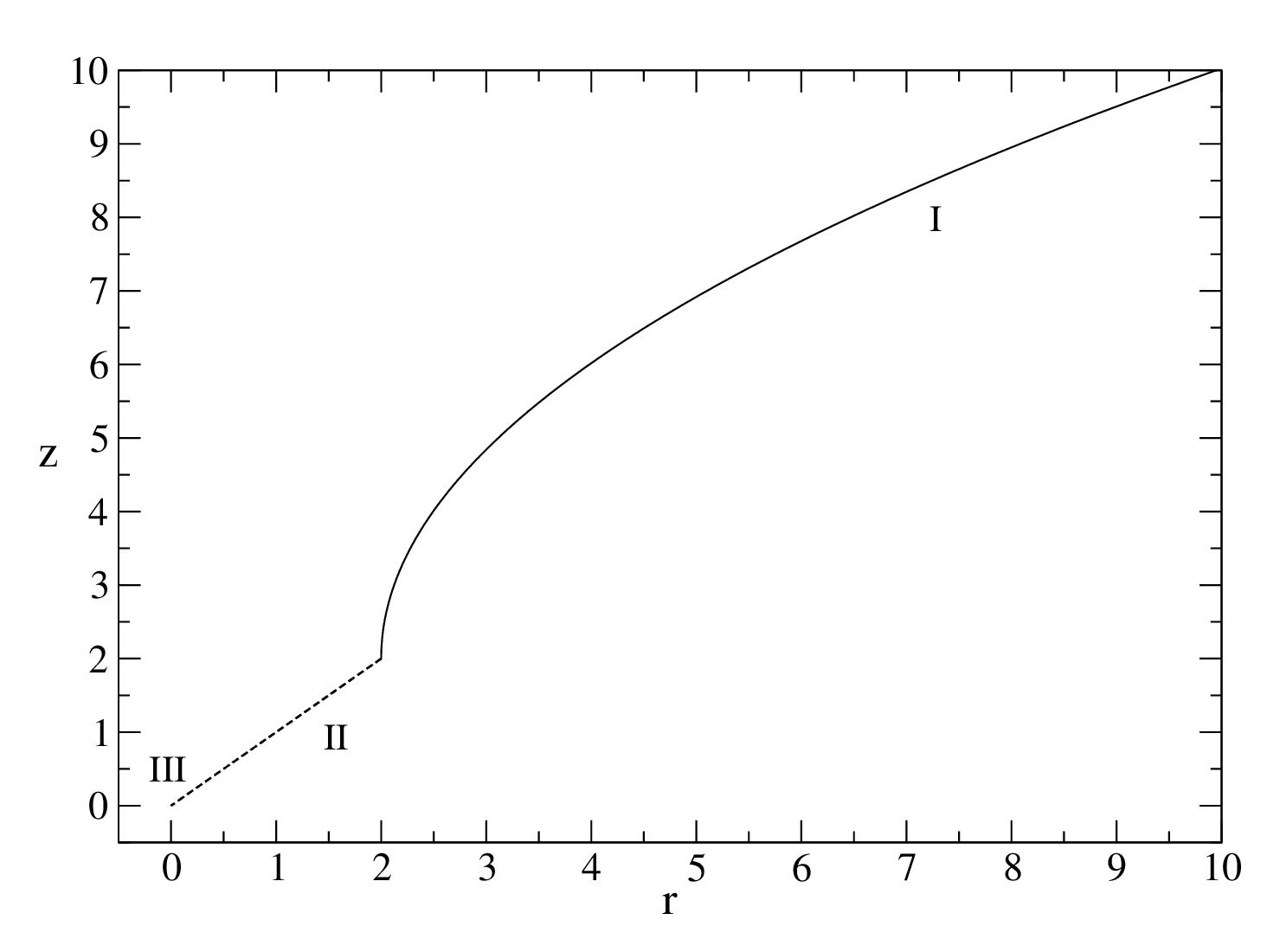}
\caption{Embedding diagram for the spacetime with time-averaged standing-wave 
scalar field of angular frequency 
$\omega=0.19$ and amplitude $\Psi_0=0.015$ at large radii 
[corresponding to the binary black hole separation $a \approx 6M$;
Eq.~(\ref{a6})].  The solid line represents embedding in Euclidean space;
the dashed line, embedding in Minkowski space.  Regions I, II and III 
are labeled on plot.}
\label{emb-fig}
\end{figure}

Figure \ref{emb-fig} shows an embedding diagram for the standing-wave spacetime:
\begin{equation}\label{emb}
\frac{dz}{dr} = \sqrt{|g_{rr}-1|}
\end{equation}
The 2-surface obtained by rotating around the vertical axis $r=0$
has the same 2-geometry as the surface $(t=\text{const},\, \theta=\pi/2)$ in
the standing-wave spacetime.  At radii $r>2$ the embedding is very nearly the
same as for a Schwarzschild black hole (cf.~Fig.~$31.5$ of \cite{MTW}).   
For $r<2$, the radial distance changes far more slowly than the areal radius 
($0<g_{rr} \ll 1$), so the embedding is performed in Minkowski space rather
than Euclidean space: the metric is $ds^2=-dz^2+dr^2+r^2d\phi^2$ rather than 
$ds^2=+dz^2+dr^2+r^2d\phi^2$.  The embedded surface asymptotes to the light cone
as $r\to0$. 

\begin{figure}[ht]
\subfigure{
\includegraphics[keepaspectratio=true,width=8.5cm,angle=0]{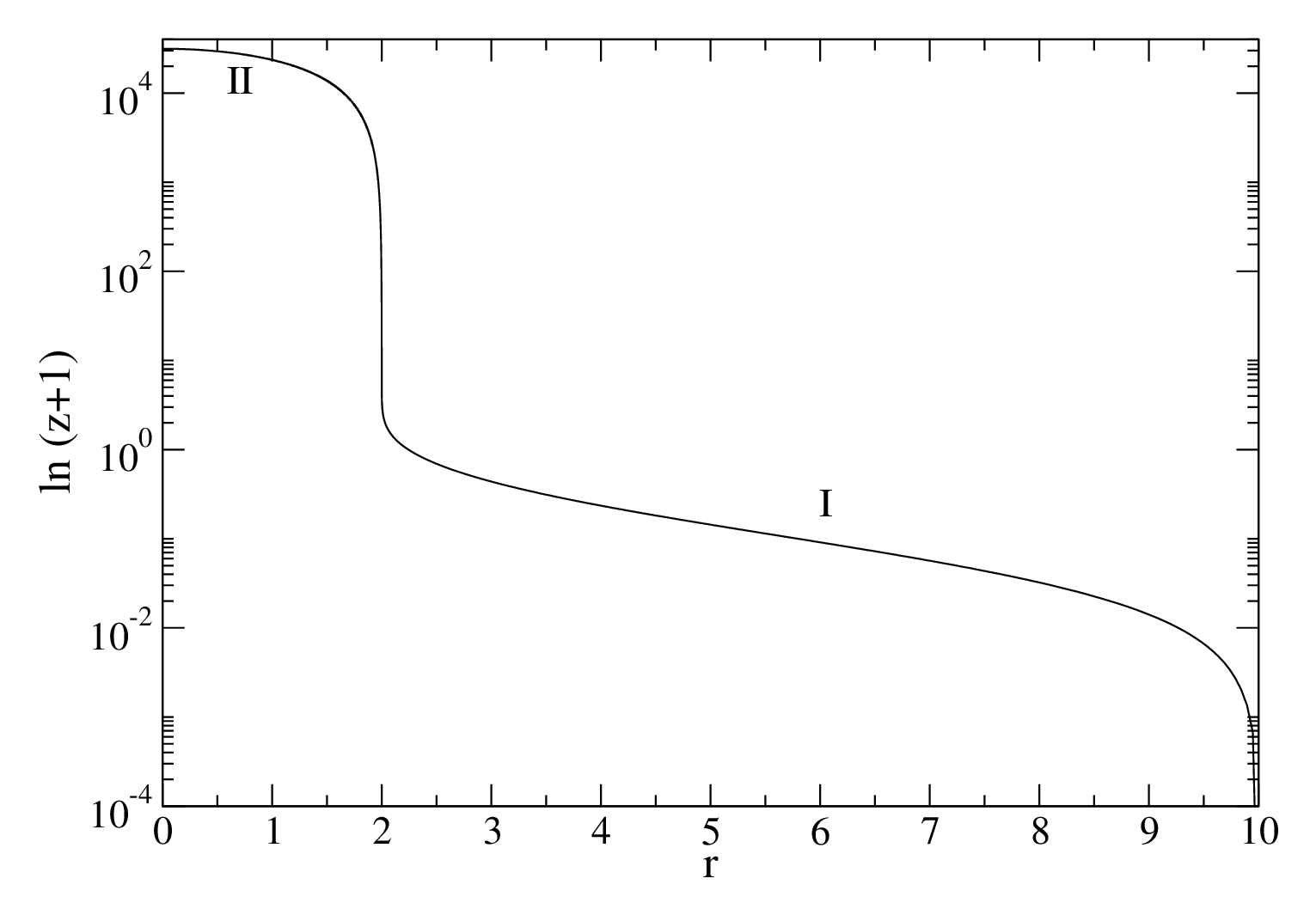}
\label{reda-fig}
}
\subfigure{
\includegraphics[keepaspectratio=true,width=8.5cm,angle=0]{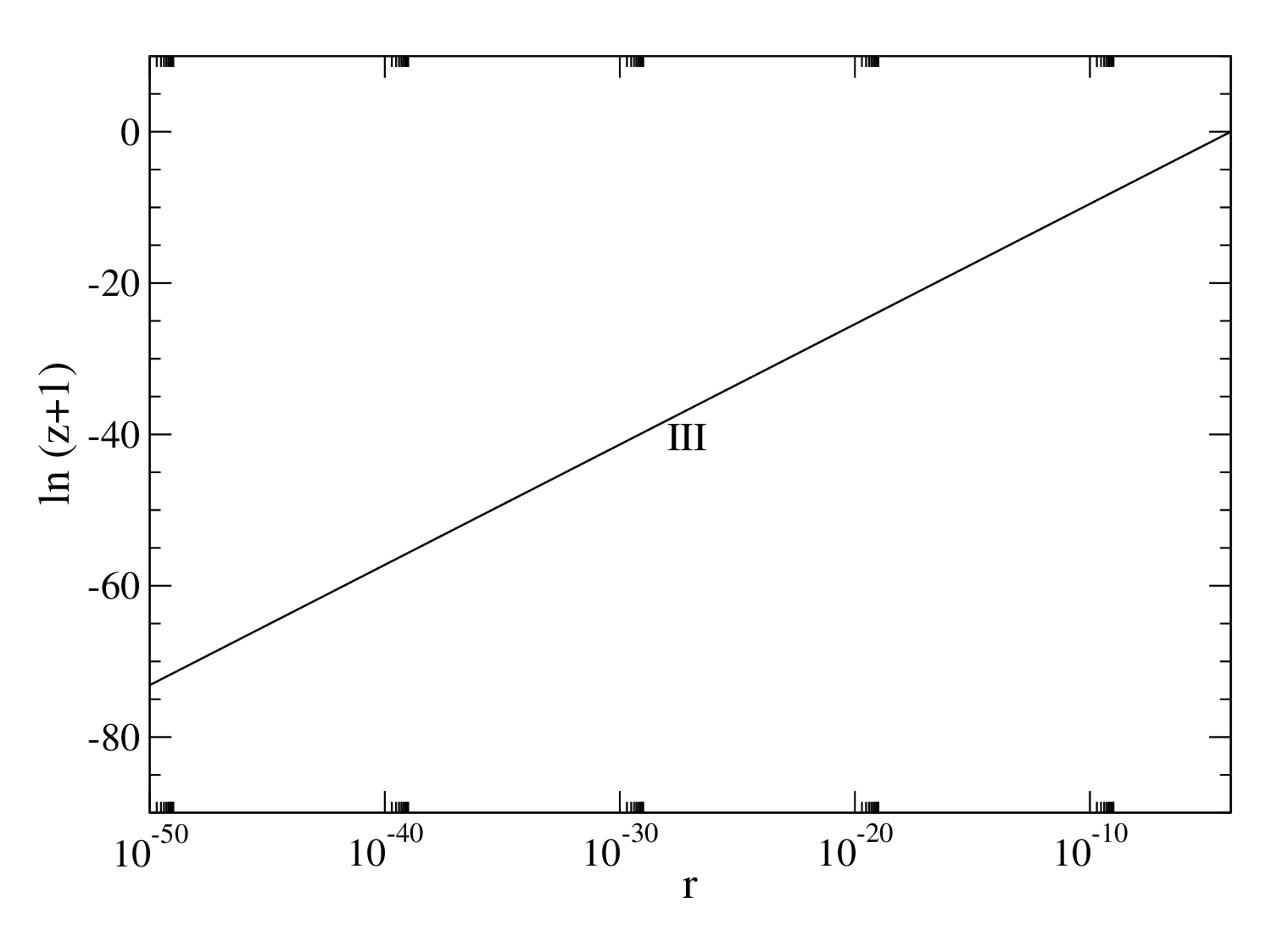}
\label{redb-fig}
}
\caption{(a) Redshift $z=\delta \lambda / \lambda$ of light emitted from
radius $r$ and received by an observer at $r=10$.  (b) Redshift for an observer 
at  $r=0.0001$.  A distant observer would see light emitted from
$r=0.0001$ redshifted by $\ln(z+1) \approx 10^5$.
These curves are drawn for the spacetime with time-averaged standing-wave 
scalar field that has angular frequency 
$\omega=0.19$ and amplitude $\Psi_0=0.015$ at large radii 
[corresponding to the binary black hole separation $a \approx 6M$;
Eq.~(\ref{a6})].}
\label{red-fig}
\end{figure}

Figure \ref{red-fig} depicts the redshift of light emitted at one radius
and received at another, greater one, as a function of the emitting radius:
\begin{equation}\label{red}
z=\sqrt{\frac{g_{tt}^{rec}}{g_{tt}^{em}}}-1
\end{equation}
Figure \ref{reda-fig} shows that, while the redshift becomes very large as 
$r \rightarrow 2$, it never becomes infinite there as it would for a 
Schwarzschild black hole.  As expected, the horizon is destroyed by 
the standing-wave scalar field, so an observer at infinity can receive 
signals from any source at $r>0$, albeit with a very large redshift for 
sources close to or inside the location ($r=2$) of the Schwarzschild 
horizon.  A blown-up view of the region $r \ll 1$ [Figure \ref{redb-fig}] 
shows that the signal emitted near the singularity may
be infinitely redshifted or blueshifted depending on the asymptotics of
the scalar field as $r \rightarrow 0$ according to
\begin{equation}\label{red3}
z=\sqrt{g_{tt}^{rec}} e^{(\alpha_0-\beta_0)/2} r^{-k^2/4+1/2} - 1 \, . 
\end{equation}

\subsubsection{Comparison of standing-wave and Schwarzschild spacetimes}
It is important to understand how the complete standing-wave spacetime 
with back reaction (we shall call this spacetime $S$) compares with
the Schwarzschild spacetime (which we shall call spacetime $D$).  We might
first try to compare the metric components in the two spacetimes as functions
of the radial coordinate $r$.  Indeed, the metric components 
$g_{\theta\theta}=r^2$ and $g_{\phi\phi}=r^2\sin^2{\theta}$ are precisely 
equal in the two spacetimes when evaluated at the same location in 
$(t, r, \theta, \phi)$ coordinates.
Furthermore, outside the effective-potential region, the perturbation 
due to the scalar field is so small that the fractional difference 
$\delta g_{\alpha\beta}/g_{\alpha\beta} \equiv 
(g^S_{\alpha\beta}-g^D_{\alpha\beta})/g^D_{\alpha\beta}$ in metric 
components $g_{tt}$ and $g_{rr}$ does not exceed $0.01\%$
for scalar-field parameters in the range of interest.
However, the metric components $g_{tt}$ and $g_{rr}$ in $S$ and $D$ can differ 
by orders of magnitude near $r=2$, inside the effective potential peak.  

The apparent mismatch between the metric components of the two spacetimes near
$r=2$ turns out to be a consequence of a poor choice of the radial 
coordinate $r$ for comparison.  A much better choice is $r^*$:  
when the coordinates $(t, r^*, \theta, \phi)$ are used for mapping between
the two spacetimes $S$ and $D$, the metric components agree extremely
well near $r=2$.

\begin{figure}[ht]
\subfigure{
\includegraphics[keepaspectratio=true,width=8.5cm,angle=0]{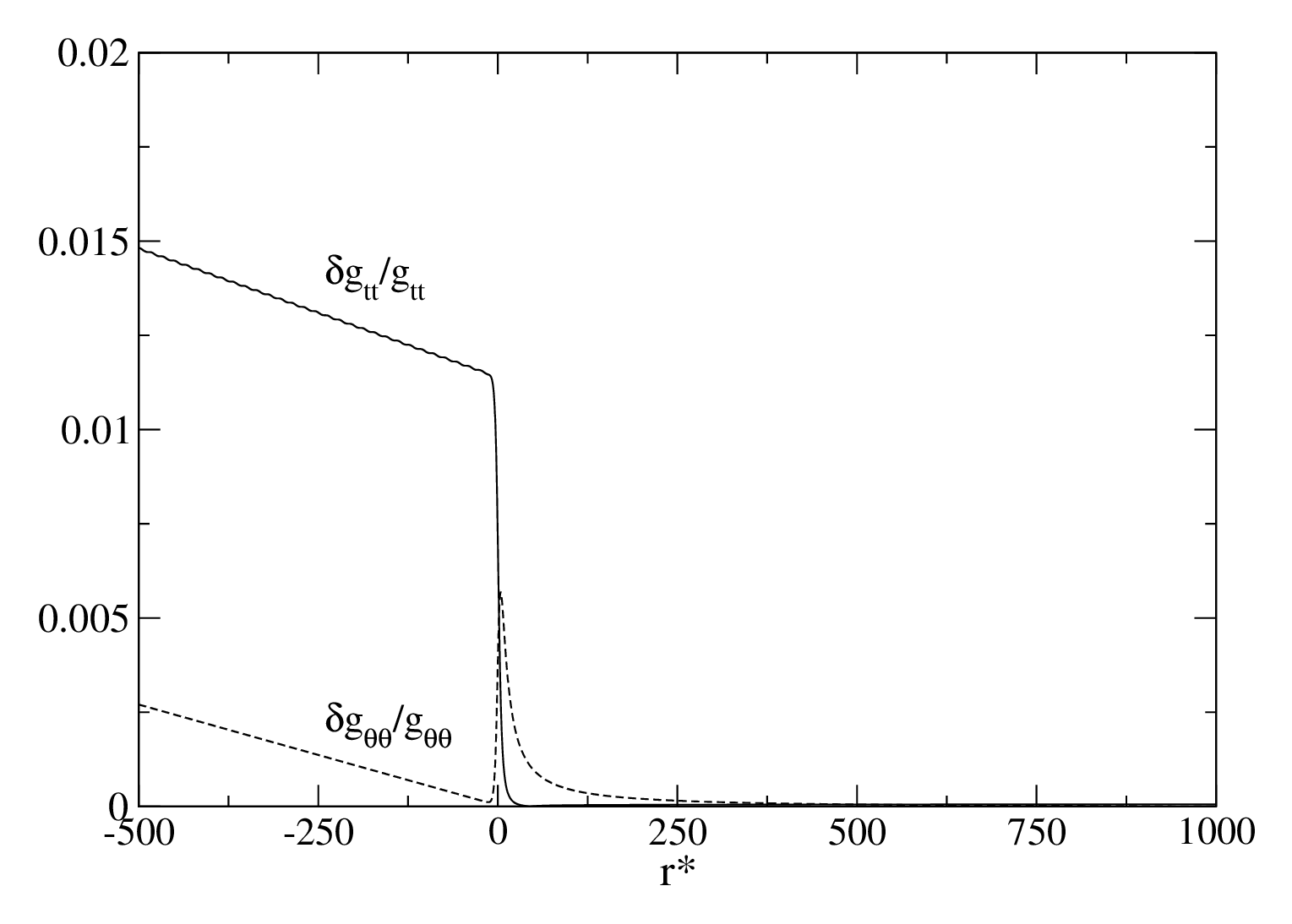}
\label{spacetime6-fig}
}
\subfigure{
\includegraphics[keepaspectratio=true,width=8.5cm,angle=0]{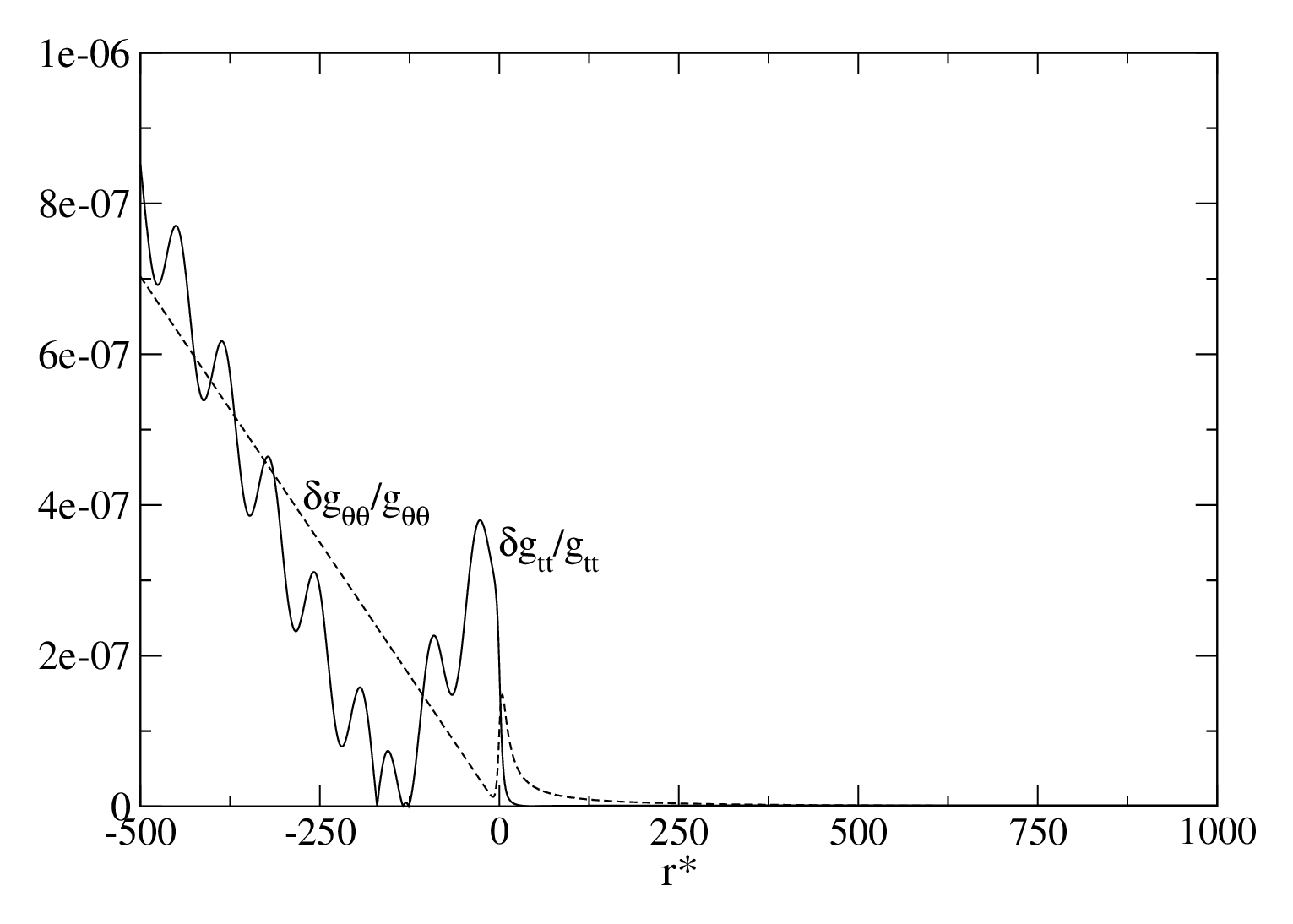}
\label{spacetime15-fig}
}
\caption{(a)Fractional differences of the metric components 
$g_{tt}=-g_{r^*r^*}$ (solid curve) and $g_{\theta\theta}$ (dashed curve) 
between Schwarzschild spacetime $D$ and standing-wave scalar field spacetime 
$S$ with scalar-wave amplitude and frequency chosen to model BBH separation 
$a \approx 6M$ [Eq.~(\ref{a6})].
(b)Same quantities plotted for scalar field parameters chosen to model
BBH separation $a \approx 15M$ [Eq.~(\ref{a15})].}
\label{spacetime-fig}
\end{figure}

The fractional differences $\delta g / g$ between 
the $g_{tt}$ and $g_{\theta\theta}$ components in $S$ and $D$ are 
plotted in Fig.~\ref{spacetime-fig} for scalar field parameters corresponding 
to binary black hole separations at the boundaries of the range of interest.  
Using $r^*$ rather than $r$ as the coordinate for comparison means that 
the $g_{\theta\theta}$ components no longer match perfectly; however, the
fractional difference introduced remains small as $r \rightarrow 2$ and 
does not exceed $0.6\%$ in the range of interest.  
The fractional differences in $g_{\phi\phi}$ are identical to those in 
$g_{\theta\theta}$ and are not plotted separately.  
The Regge-Wheeler tortoise coordinate $r^*$ 
[Eq.~(\ref{rRW})] and its generalization [Eq.~(\ref{Dr})] were defined
so that $g_{r^*r^*} \equiv -g_{tt}$ in both spacetimes $S$ and $D$; therefore,
the fractional differences in the values of $g_{r^*r^*}$ in $S$ and $D$ are 
the same as the fractional differences in $g_{tt}$.

As Fig.~\ref{spacetime-fig} shows, the fractional differences in the metrics 
are $\lesssim 0.02$ down to values of $r^* \sim -1000$, a location so deep 
inside the peak of the effective potential that it contains  at least $20$ 
near-horizon oscillations of the scalar field for frequencies and 
amplitudes in the BBH separation range of interest.  
Perhaps a more impressive way to state this is that 
in the $(t, r^*, \theta, \phi)$  coordinate system, metric 
components of $g^S$ and  $g^D$ match to an accuracy of $2\%$ 
for all relevant scalar-field parameters down to the Schwarzschild radius 
$r^D - 2 < 10^{-100}$.  

The fractional differences between the coefficients of the metrics $g^S$ and 
$g^D$ continue to grow approximately linearly in $r^*$ deep inside the
effective potential and reach $10\%$ at the Schwarzschild radius
$r^D - 2 \sim 10^{-3000}$, or approximately 500 scalar-field 
oscillations inside the effective-potential peak for scalar field parameters
corresponding to BBH separation $a \approx 6 M$.

\section{\label{sec:down}Downgoing scalar field}

Having discussed, in Sec.~\ref{sec:sw}, the standing-wave scalar-field 
spacetime that modeled the stationary BBH approximation, we now turn to a
scalar-field spacetime that  serves as a model for the physical BBH spacetime
with downgoing gravitational waves at the black-hole horizons:
a nearly Schwarzschild spacetime with spherically symmetric scalar waves that 
are purely downgoing at $r=2$.  

For a perturbative analysis of downgoing scalar waves in Schwarzschild, the 
ingoing, null Eddington-Finkelstein time coordinate $v=t+r^*$ is more 
appropriate than the standard Schwarzschild time coordinate.  Let us suppose 
that by the time $v=0$ a total mass-energy $M_0 = m(v=0)$ is located within 
the horizon $r=2$.  We are not particularly interested in how this mass 
accumulated there or how the scalar field behaved in the past; we are only 
interested in the times immediately following $v=0$, and we let the scalar 
waves be purely downgoing and monochromatic at the horizon for $v>0$.  Then for 
$v>0$ radiation is falling into the black hole at a nearly constant rate, 
corresponding to the energy density in the scalar field 
$dm/dv \approx \Psi_0^2 \omega^2 / 2$, with some small oscillations 
on top of the linear increase in mass.  This is very similar to the Vaidya
solution and, indeed, the Vaidya metric will be seen to describe the 
spacetime of the downgoing scalar-field solution:
\begin{equation}\label{vaidya}
ds^2 = -\left[1-\frac{2m(v)}{r}\right] dv^2 + 2dv dr + r^2 d\Omega^2 \, .
\end{equation}

Near the horizon, $\Phi=(1/r)\cos{\omega v}$ is a purely downgoing 
solution to the wave equation (\ref{wave}).  The only non-zero term of the 
Ricci tensor in Vaidya coordinates is $R_{vv}=(2/r^2) m'(v)$, where
$'$ denotes the derivative with respect to $v$.  The Einstein 
equations (\ref{Ricci}) at $r=2$ say:
\begin{equation}\label{Riccidown}
R_{vv}=\frac{2 m'(v)}{4} = 2 \Phi_{,v} \Phi_{,v} = 
	\frac {2 \Psi_0^2 \omega^2 \sin^2{\omega v}}{4} \, .
\end{equation}

Equation (\ref{Riccidown}) is trivially integrated to obtain:
\begin{equation}\label{mdown}
m(v) = M_0 + \frac{\Psi_0^2 \omega^2}{2} v - 
	\frac{\Psi_0^2 \omega \sin{2\omega v}}{4} \, .
\end{equation}
The black-hole mass grows linearly in $v$ at the rate $\Psi_0^2 \omega^2/2$
with a tiny superimposed oscillatory component.  The black hole retains a
smooth, non-singular future horizon.

The scalar field is purely downgoing at the horizon and approximately downgoing 
everywhere inside the Schwarzschild effective-potential peak.  Outside the 
effective-potential peak there is both a downgoing scalar field and an upgoing 
one, reflected off the potential.  Since for small $v$ the metric is nearly 
Schwarzschild [the constant Schwarzschild mass $M$ is replaced by the $m(v)$ 
of Eq.~(\ref{mdown})], the scalar field everywhere is given to a high accuracy 
by a solution to the wave equation in the Schwarzschild background subject to 
the purely downgoing boundary condition at the horizon.  (Of course, very far 
from the horizon the energy contained in the intervening scalar field will act 
as an additional mass, but we are not interested in this region for our model 
problem.)

\section{\label{sec:reconstruct}Reconstruction of downgoing scalar field 
from standing-wave scalar field}

We turn now to our scalar-wave version of adding a radiation-reaction
field to a standing-wave spacetime to obtain a physical spacetime with
downgoing waves at horizons and outgoing waves at infinity.  For this
procedure there is a substantial difference between the BBH problem and our
model problem.

In the true BBH problem, the periodic standing wave (SW) solution is sourced
by the black holes and corresponds to the $\frac{1}{2} \text{Retarded} + 
\frac{1}{2} \text{Advanced}$ solution of the Green's function problem.  In 
this case we add the non-sourced $\frac{1}{2} \text{Retarded} - 
\frac{1}{2} \text{Advanced}$ radiation reaction (RR) solution of the linearized
Einstein equations in the SW spacetime to get an approximation to the 
physical retarded solution \cite{private}.  At infinity, where the SW 
field is $\frac{1}{2} \text{Outgoing} + \frac{1}{2} \text{Ingoing}$, the
boundary condition for the RR field should be set to
$\frac{1}{2} \text{Outgoing} - \frac{1}{2} \text{Ingoing}$, so
that their sum contains only physical outgoing waves, and similarly at the 
horizons the RR field will be 
$\frac{1}{2} \text{Downgoing} - \frac{1}{2} \text{Upgoing}$.  Adding this 
RR field to the $\frac{1}{2} \text{Downgoing} + 
\frac{1}{2} \text{Upgoing}$ standing waves would yield gravitational waves
that are downgoing at the expected horizon locations, conforming
to the expected behavior in physical black-hole spacetimes.  (We do not
expect the stress-energy tensor of the sum of SW and RR waves to precisely 
match the Einstein tensor of the SW spacetime because, of course, gravitational 
theory is not linear; however, it is likely that "effective linearity" holds 
in the sense defined by Price \cite{Price} for the non-tidally-locked case as 
well as for the tidally-locked case.  In a future paper we intend to explore
this issue with a model that more closely resembles the BBH problem.)

The scalar-field model we are currently analyzing is not sourced:
the wave equation (\ref{wave}) we used to compute the SW solution 
is homogeneous.  There is then no perturbative homogeneous
solution that is simultaneously $\frac{1}{2} \text{Outgoing} - 
\frac{1}{2} \text{Ingoing}$ at infinity and $\frac{1}{2} \text{Downgoing} 
- \frac{1}{2} \text{Upgoing}$ at the expected horizon location.  
Since at the outer boundary the problem is obviously linear for 
sufficiently weak scalar fields, it is easy to reconstruct the outgoing 
solution from the SW solution there: we simply double the outgoing 
component of the SW solution.  The interesting case lies in the extraction 
of a downgoing solution near $r \approx 2$.  We attempt to reconstruct the
downgoing scalar field from the SW scalar field near the expected horizon
by adding to the SW field a perturbative RR field that is
$\frac{1}{2} \text{Downgoing} - \frac{1}{2} \text{Upgoing}$ at 
$r \approx 2$.

As in Sec.~\ref{sec:sw-nonlin}, let $S$ denote the spacetime of the complete 
standing-wave solution with back reaction.  As discussed in the previous
section, the spacetime of the downgoing scalar field is approximated to
sufficient accuracy for our purposes by the Schwarzschild spacetime $D$.

The complete SW scalar field is a solution to the wave equation in 
spacetime $S$ (in our simplified treatment of the problem, spacetime $S$ 
actually corresponds to the time-averaged solution, i.e., one in which we 
ignore the oscillatory components of the metric). The RR field is a solution 
to the same wave equation in $S$ in our model.  The ``reconstructed'' downgoing 
field is, therefore, the downgoing solution to the wave equation in $S$.  
We want to compare this to the ``true'' downgoing field, which is the downgoing 
solution to the wave equation in $D$, i.e., in Schwarzschild. 
 
\begin{figure}[tbh!]
\subfigure{
\includegraphics[keepaspectratio=true,width=8.5cm,angle=0]{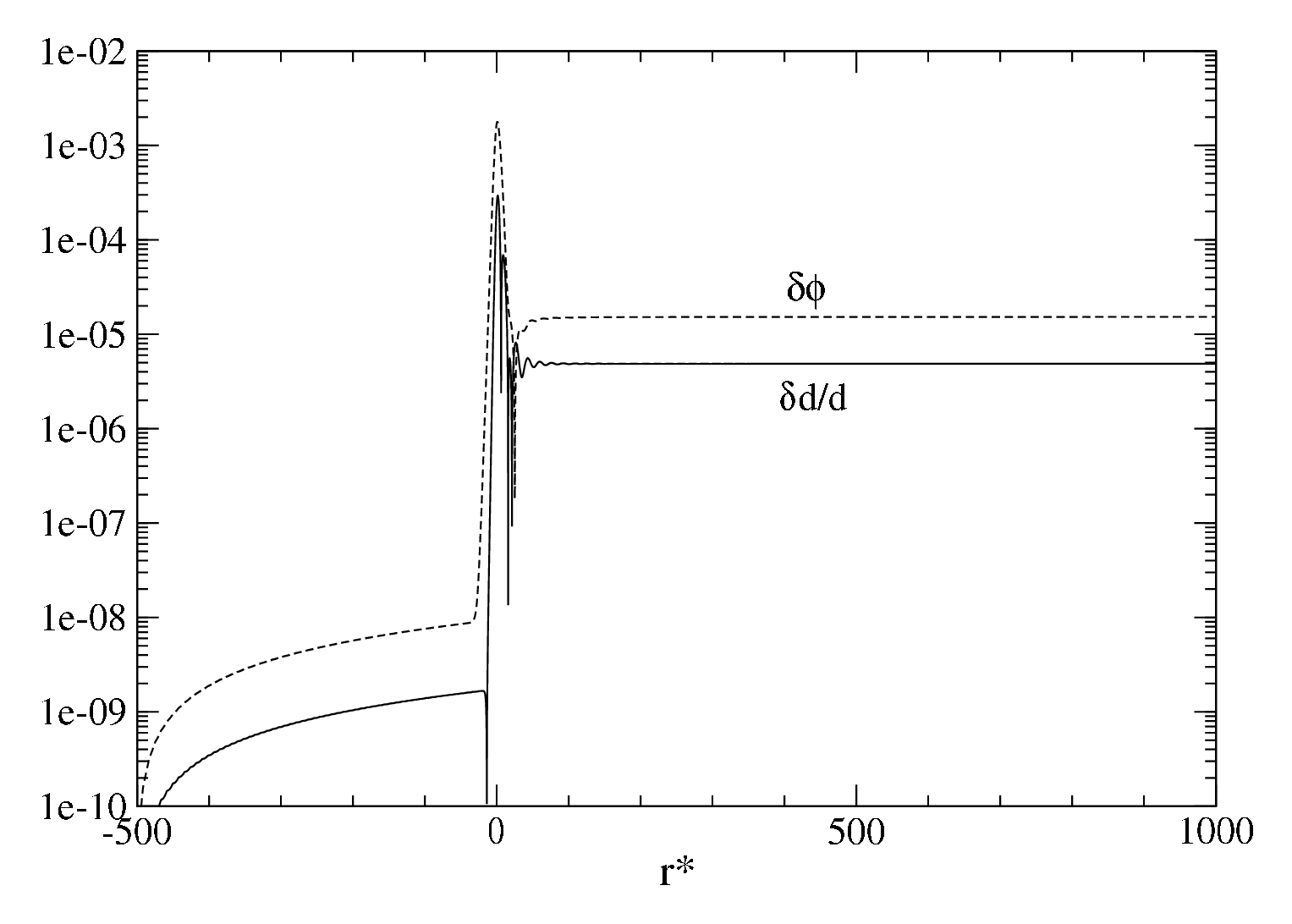}
\label{recona-fig}
}
\subfigure{
\includegraphics[keepaspectratio=true,width=8.5cm,angle=0]{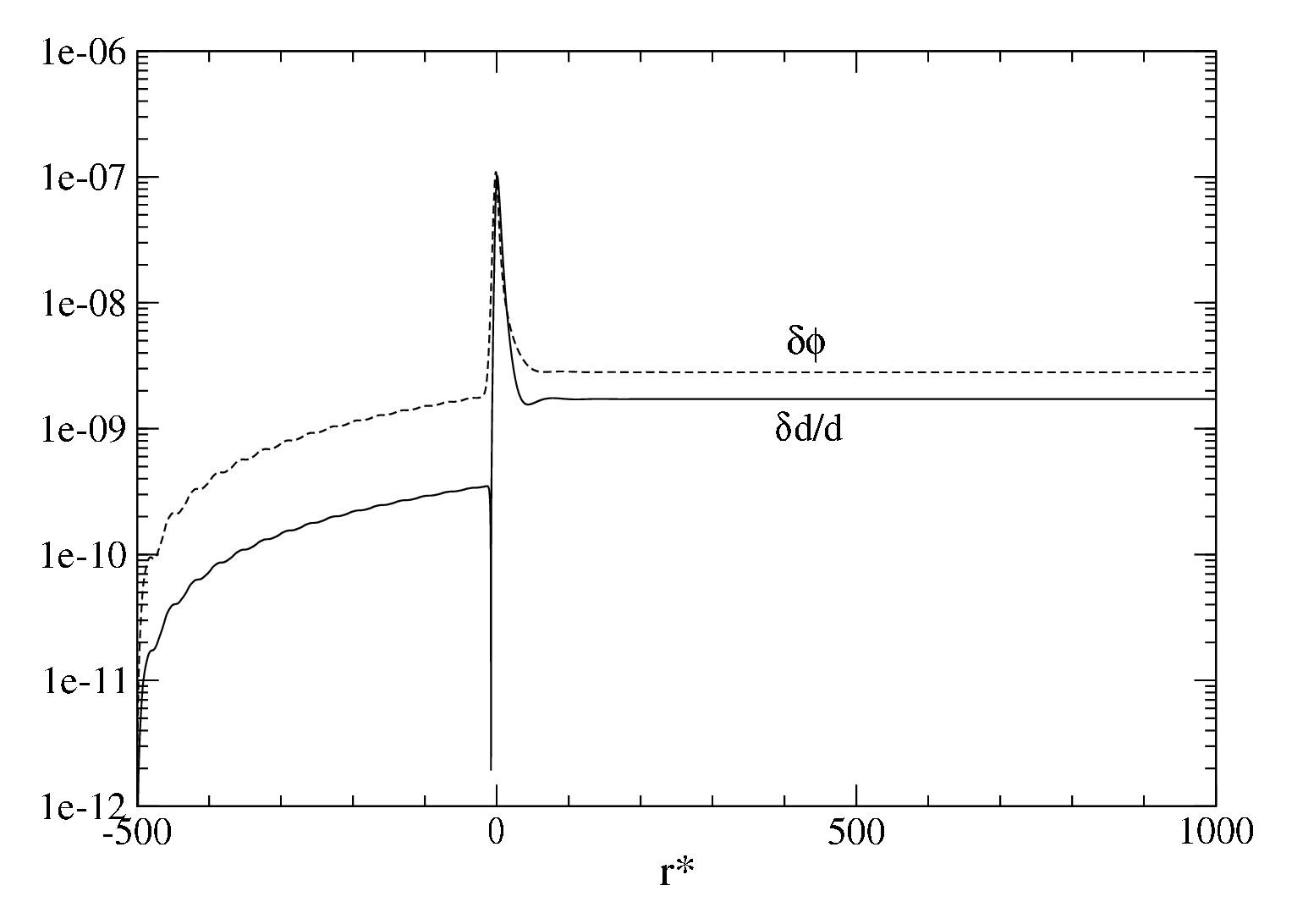}
\label{reconb-fig}
}
\caption{(a) The fractional difference in the amplitudes of the reconstructed 
scalar field and downgoing scalar field 
$\delta d / d \equiv (d^{SW+RR}-d^{down})/d^{down}$
(solid curve) and the phase difference between the 
two fields $\delta \phi_d = \phi_d^{SW+RR}-\phi_d^{down}$ (dashed curve),
plotted vs. $r^*$.  Scalar-wave amplitude and frequency chosen to model BBH
separation $a \approx 6M$.
(b) Same quantities plotted for scalar-field parameters chosen to model 
BBH separation $a \approx 15M$.}
\label{recon-fig}
\end{figure} 
 
In the region between the expected horizon location $r=2$ and the inner edge 
of the peak of the effective potential, the wave equation (\ref{wave}) is 
dominated by 
\begin{equation}\label{appwave}
\frac{d^2\Psi}{d{r^*}^2} \approx -\omega^2 \Psi 
\end{equation}
in both spacetimes $S$ and $D$.  Hence, the solution to the wave equation
will be oscillatory in $r^*$ with frequency $\omega$, which makes
sense on physical grounds, since ingoing light cones are 
$t+r^*=\text{constant}$ in both $S$ and $D$.  Moreover, as discussed 
in  Sec.~\ref{sec:sw-nonlin}, the metrics of the two spacetimes
are nearly the same in the $r^*$ coordinate, i.e., $g^S(r^*) \approx g^D(r^*)$. 
This suggests that to get the scalar wave phasing to agree, we need
to map between the two spacetimes using the $r^*$ radial coordinate.

We set the boundary conditions for both the RR field in $S$ and the downgoing
field in $D$ at a negative value of $r^*$ chosen so that the fields are
at least a few wavelengths inside the effective potential, and so that 
$r^\text{S}(r^*)$ is very close to $r^\text{S}=2$ (it might actually be 
slightly inside $r=2$).  The $\text{SW}+\text{RR}$ and downgoing scalar fields 
will match by construction at the point where the initial conditions are set.  
We will integrate both solutions toward larger $r^*$ and compare the quality of
the match between the two fields.

For the purposes of comparing the scalar fields in the two spacetimes,
we separate the complex scalar field $\Psi(r^*)$ [the spatial factor of 
the complete field $\Phi(r,t)=\Re[\Psi(r^*) e^{-i\omega t}]/r$, 
cf.~Eq.~(\ref{Phi})] into upgoing and downgoing components.  We define
the amplitudes and phases of the upgoing and downgoing fields as follows (see
below for motivation):
\begin{subequations}\label{PsiDU}
\begin{eqnarray}
u&\equiv&\frac{1}{2\omega} \left| \frac{d\Psi}{dr^*}+i\omega\Psi\ \right| ;
\\
d&\equiv&\frac{1}{2\omega} \left| \frac{d\Psi}{dr^*}-i\omega\Psi\ \right| ;
\\
e^{i\phi_u}&\equiv&\frac{1}{2i\omega u} 
	\left( \frac{d\Psi}{dr^*}+i\omega\Psi\ \right) ;
\\
e^{i\phi_d}&\equiv&\frac{1}{-2i\omega d} 
	\left( \frac{d\Psi}{dr^*}-i\omega\Psi\ \right) \, .
\end{eqnarray}
\end{subequations}

To motivate these definitions we consider the
geometric optics limit, where the wave phase evolves much 
faster than the amplitude.  In this limit,
the downgoing component of the scalar field 
$\Psi_d \propto e^{-i\omega r^*}$ separates unambiguously
from the upgoing component $\Psi_u \propto e^{i\omega r^*}$:
\begin{subequations}\label{PsiDUmot}
\begin{equation}\label{Psicomp}
\Psi(r^*) =  u e^{i \phi_u} + d e^{i \phi_d} \, ,
\end{equation}
where we use the standard approximations
\begin{eqnarray}\label{phiu}
\frac{d \phi_u}{dr^*} &\cong& \omega \gg \left| \frac{d u}{dr^*}\right| ,
\\
\label{phid}
- \frac{d \phi_d}{dr^*} &\cong& \omega \gg \left|\frac{d d}{dr^*}\right| \, .
\end{eqnarray}
\end{subequations}
Inverting Eq.~(\ref{Psicomp}) with these approximations yields the definitions
(\ref{PsiDU}).
Although the geometric optics approximations break down in the region of 
the effective potential, and the separation of the scalar waves into 
upgoing and downgoing components becomes ambiguous there because the wave
speed is ill-determined outside the short-wavelength limit, expressions
(\ref{PsiDU}) are adequate for comparing scalar fields in our region
of interest.  

In Fig.~\ref{recon-fig} we show the fractional difference 
$\delta d / d \equiv (d^{SW+RR}-d^{down})/d^{down}$ in the amplitude of 
the downgoing components of the reconstructed $\text{SW}+\text{RR}$ waves 
and the downgoing waves along with the phase difference 
$\delta \phi_d = \phi_d^{SW+RR}-\phi_d^{down}$.
The two plots represent the endpoints of the range of relevant BBH 
separations: $a\approx 6M$ in Fig.~\ref{recona-fig} and 
$a\approx 15M$ in Fig.~\ref{reconb-fig}.  Only the downgoing 
amplitude $d$ and downgoing phase $\phi_d$ are plotted.  The upgoing field 
components are zero to numerical precision inside the effective potential 
and the differences between the reflected upgoing components of the 
``reconstructed'' and ``true'' downgoing fields outside the effective-potential 
peak are similar to the differences between the downgoing field components 
there, $\delta u/u \sim \delta_d/d$ and $\delta \phi_u \sim \delta \phi_d$.

The amplitudes and phases of the ``true'' downgoing field and the 
``reconstructed'' downgoing field match to within one part in ten million
from the location where the initial conditions are set 
(several scalar-field oscillations inside the effective potential) 
to the inner edge of the effective-potential peak for all BBH separations in 
the range of interest.  Near the effective-potential peak the 
fractional difference in the amplitudes does not exceed $0.03\%$ and the phase 
difference is less than $0.002$.  Outside the effective potential, the 
fractional difference in the amplitudes is $5$ parts per million and the phase 
difference is less than $0.00002$ for the smallest BBH separations in the range
of interest.  

We also compared the ``reconstructed'' and ``true'' downgoing fields 
very deep inside the effective potential when the field-matching initial
conditions are set about 10 scalar-field oscillations inside the 
effective-potential peak.  In this case, the amplitudes of the two fields are 
equal to within numerical precision and the phase difference does not exceed 
$3 \times 10^{-7}$ down to $500$ scalar-field oscillations inside the 
effective-potential peak.  The fields begin to disagree significantly only 
once the naked singularity is approached in the spacetime $S$, at
$r^S(r^*) \lesssim 0.2$ 
\footnote{It might seem odd that the fields continue to match far deeper 
(at far more negative $r^*$) than the metrics, which begin to disagree by 
$10\%$ at $500$ scalar-field oscillations inside the
effective-potential peak.  The reason is that in the wave equation, the metric 
enters only into the effective-potential piece [see Eq.~(\ref{DPsi})], 
which is so tiny throughout the region $0.2 \lesssim r^S \lesssim 2$ that even 
significant deviations of the standing-wave spacetime metric $g^S$ from the
Schwarzschild metric do not affect the behavior of the scalar field.}.

\begin{acknowledgments}
I am very grateful to my advisor, Kip S. Thorne, for posing this research 
project and for his extremely patient guidance.  Jonathan Gair, Lee Lindblom, 
Robert Owen, Frans Pretorius and Mark A. Scheel of the TAPIR group at Caltech 
made many helpful suggestions, as did Ronald J. Adler and Alexander S. 
Silbergleit of the GP-B theory seminar at Stanford.  I was partially supported 
by NSF Grant PHY-0099568 and NASA Grant NAG5-12834.
\end{acknowledgments}


\begin{thebibliography}{99}
\bibitem{GravRad}
K. S. Thorne, in 300 Years of Gravitation, edited by S. Hawking 
and W. Israel (Cambridge University Press, 1987)
\bibitem{Brady}
P. R. Brady, J. D. E. Creighton, and K. S. Thorne, Phys. Rev. D
\textbf{58}, 61501 (1998)
\bibitem{Detweiler}
S. Detweiler, Phys. Rev. D \textbf{50}, 4929 (1994)
\bibitem{Price}
R. Price, Class Quant. Grav. \textbf{21}, S281-S293 (2004);
Z. Andrade \textit{et al.}, Phys. Rev. D \textbf{70}, 064001 (2004)
\bibitem{private}
K. S. Thorne, unpublished notes, 2002
\bibitem{Wald}
H. Friedrich, I. R\'acz, and R. M. Wald, Commun. Math. Phys.
\textbf{204}, 691 (1999)
\bibitem{MTW}
C. W. Misner, K. S. Thorne, and J. A. Wheeler, \textit{Gravitation} 
(Freeman, San Francisco, 1973)
\bibitem{NF}
V. P. Frolov and I. D. Novikov, \textit{Black Hole Physics}, 
Fundamental Theories of Physics (Kluwer, Dordrecht, 1998)
\bibitem{Poisson}
E. Poisson and M. Sasaki, Phys. Rev. D \textbf{51}, 5753 (1995)
\bibitem{Hua}
H. Fang and G. Lovelace, in preparation
\bibitem{RW}
T. Regge and J. A. Wheeler, Phys. Rev. \textbf{108}, 1063 (1957)
\end{thebibliography}
\end{document}